\documentclass[prb,twocolumn,showpacs,preprintnumbers]{revtex4}
\usepackage{graphicx}
\usepackage{dcolumn}

\newcommand{\beq}{\begin{equation}}
\newcommand{\eeq}{\end{equation}}
\newcommand{\beqa}{\begin{eqnarray}}
\newcommand{\eeqa}{\end{eqnarray}}

\begin{document}

\title{Cluster Dynamical Mean-Field Theory of the density-driven Mott
transition in the one-dimensional Hubbard model}
\author{M. Capone$^1$, M. Civelli$^2$,  S. S.  Kancharla{$^2$,$^3$}, 
C. Castellani {$^4$}, and G. Kotliar$^2$}
\affiliation{ $^1$ Enrico Fermi Center, Rome, Italy}
\affiliation{$^2$ Physics Department and Center for Materials
Theory, Rutgers University, Piscataway NJ USA}
 \affiliation {$^3$ Departement de physique and
Regroupement quebecois sur les materiaux de pointe, Universit\'e de
Sherbrooke, Sherbrooke, Quebec J1K 2R1, Canada}

\affiliation{$^4$ Physics Department, University of Rome ``La
Sapienza'', and INFM Center for Statistical Mechanics and
Complexity, Piazzale A. Moro 5, I-00185, Rome, Italy}

\begin{abstract}
The one-dimensional Hubbard model is investigated by means of two different
cluster schemes suited to introduce short-range spatial correlations beyond
the single-site Dynamical Mean-Field Theory, namely the Cellular Dynamical
Mean-Field Theory, which does not impose the lattice symmetries, and 
its periodized version in which translational symmetry is recovered. 
It is shown that both cluster schemes are able to
describe with extreme accuracy the evolution of the density as a function of
the chemical potential from the Mott insulator to the metallic state. Using
exact diagonalization to solve the cluster impurity model, we discuss the
role of the truncation of the Hilbert space of the bath, and propose an
algorithm that gives higher weights to the low frequency hybridization matrix
elements and improves the speed of the convergence of the algorithm.
\end{abstract}

\date{\today}
\maketitle

Strongly correlated electronic systems and the models describing them
represent a formidable challenge for theorists. An important advance in this
field has been achieved through the development of the Dynamical Mean-Field
Theory (DMFT),\cite{revdmft} in its single-site version.
DMFT is  a non perturbative
approach which fully retains local quantum dynamics, but simplifies the
spatial dependence of the correlation functions to make the problem
tractable. The lattice problem is therefore mapped onto a dynamical local
problem, and consequently onto an impurity model subject to a
self-consistency relation. The DMFT proved its power providing the first
unified scenario for the longstanding problem of the Mott transition in
the Hubbard model, and
completely characterizing the peculiar criticality associated with this
transition. Numerous theoretical predictions of this approach have been
verified experimentally. \cite{science} The combination of single-site DMFT
with electronic structure methods has given new insights into the physical
properties of many correlated materials. The main limitation of DMFT is the
neglect of spatial correlation in the one-electron spectra, which makes
impossible to treat phases with definite spatial ordering such as d-wave
superconductivity. Non local effects, such as short-range order,
can also be very important in describing the metal-insulator transition in
materials such as Ti$_2$O$_3$. \cite{poteryaev}

There is therefore a strong motivation for developing extensions
of single-site DMFT. Several schemes have been proposed, and they
introduce short-range correlations by replacing the single
impurity model with a cluster-impurity, containing $N_{c}$ sites
with a given spatial arrangement. The Dynamical Cluster Approximation\cite{dca}
consists in a systematic inclusion of a few lattice momenta, which would
correspond to the lattice momenta of an $N_{c}$-site cluster.
From the impurity model point of view, this approach requires
periodic boundary conditions on the cluster.\cite{dca} A
different approach uses a continuous interpolation of the 
self-energy in momentum space in the self consistency condition.
\cite{lichtenstein} An alternative cluster method, the
Cellular Dynamical Mean-Field Theory (CDMFT), has been proposed in
Refs. \onlinecite{cdmft,bk}, closely following the spirit of the
DMFT. In this approach, the cluster has open boundary conditions,
and a 'local' theory for the cluster degrees of freedom is
obtained through the cavity method. The basic approximation is to
assume that the dynamical field experienced by the cluster is
Gaussian. A modification of the CDMFT scheme, the
Periodized-CDMFT (PCDMFT),\cite{bpk} enforces the lattice
periodicity at every iteration.  Both PCMDFT and CDMFT have been
proved to give causal spectral functions. \cite{bpk}

In this work we focus on the CDMFT and PCDMFT and study their performance in
the one-dimensional Hubbard model, which is the worst case scenario for
every mean-field theory. The ability of CDMFT to capture the physics of the
Mott insulating state in one dimension has been already demonstrated in Ref.
\onlinecite{venky}, where the half-filled Hubbard chain was
correctly found to be an insulator for every value of $U$, with a
Mott-Hubbard gap which closely follows the Bethe Ansatz (BA)
exact solution.\cite{liebewu}
Here we extend the analysis of Ref. \onlinecite{venky} to the
doped metallic state, and compare CDMFT and PCDMFT to the BA
results for the whole density range and different values of the
interaction strength. This is a much more stringent test of the
method, since metallic states, with gapless excitations, are
more difficult to describe with a reduced number of sites than
insulators.  One goal of the investigation is to compare the
performance of CDMFT and PCDMFT, with an eye to future
applications. As in Ref. \onlinecite{venky}, we use exact
diagonalization to solve the cluster-impurity model. A non
trivial issue is how to discretize the bath to implement
the solution of the cluster impurity model by exact
diagonalization. 
This problem was considered in Refs.\onlinecite{caffarel,exactdiag}
in the context of single-site DMFT, and, more recently,
by Potthoff in the context of
cluster methods using a functional technique.\cite{potthoff}
Here we find that a
simple modification of the notion of distance in the algorithm of
Ref. \onlinecite{caffarel} improves dramatically the quality and the
convergence of the results.
The paper is organized as follows: Sec. II introduces the model and our
theoretical tools, Sec. III presents our results and Sec. IV is dedicated
to the conclusions.
\section{model and method}
We consider the one-dimensional Hubbard model
\begin{equation}
H = -t\sum_{\langle i,j\rangle,\sigma} (c^{\dagger}_{i,\sigma} c_{j,\sigma} +
h.c.) + U \sum_i n_{i\uparrow}n_{i\downarrow} -\mu\sum_i n_i,
\label{hamiltonian}
\end{equation}
where $c_{i,\sigma}$ ($c^{\dagger}_{i,\sigma}$) are destruction (creation)
operators for electrons of spin $\sigma$, $n_{i\sigma} = c^{\dagger}_{i\sigma}
c_{i\sigma}$ is the density of 
$\sigma$-spin electrons, $t$ is the hopping amplitude, $U$ is the on-site 
repulsion and $\mu$ the chemical potential.

Within cluster DMFT methods an effective action for the cluster degrees of
freedom is defined as
\begin{eqnarray}
S^{eff}&=&\int_{-\beta}^{\beta}d\tau d\tau^{\prime} \sum_{\mu\nu\sigma}
c^{\dagger}_{\mu\sigma}(\tau) {\mathcal{G}_{\mu\nu\sigma}^{-1}(\tau-
\tau^{\prime})} c_{\nu\sigma} (\tau^{\prime})+  \nonumber \\
&+& \int_{-\beta}^{\beta} d\tau \sum_{\mu=1}^{N_c} U n_{\mu\uparrow}(\tau)
n_{\mu\downarrow}(\tau),  \label{seff}
\end{eqnarray}
where ${{\mathcal{G}}_{\mu\nu\sigma}^{-1}}$ is a Weiss dynamical field and
$\mu,\nu=1,\ldots N_c$ are indices of sites in the cluster. 
The action for the cluster degrees of freedom can be derived, in close
analogy with DMFT, by means of the cavity construction, i.e.,by integrating out
the degrees of freedom of all the fermions except the ones in the cluster,
and assuming that the dynamical field due to the other sites is Gaussian.\cite{cdmft}
The cluster has therefore naturally open boundary conditions, as opposed
to the periodic boundary conditions of the Dynamical Cluster Approximation
\cite{dca}, and breaks the translational symmetry.
In practical implementations of this approach, 
the initial dynamical Weiss field is specified by the choice of a given
${{\mathcal{G}}_{\mu\nu\sigma}^{-1}}$. 
The effective action is then solved with some technique, providing
the cluster Green's function  $G_{\mu\nu\sigma}(\tau) = -\langle T_{\tau}
c_{\mu\sigma}(\tau) c^{\dagger}_{\nu\sigma} \rangle$. 
Consequently, the cluster
self-energy is obtained (dropping the spin index to simplify the notations) as
\begin{equation}
\label{sigma}
\Sigma_{\mu \nu}^{c}(i\omega_{n})=\mathcal{G}_{\mu \nu}^{-1}(i\omega_{n})-
G_{\mu \nu}^{-1}(i\omega_{n}).  \label{selfenergy}
\end{equation}

Just like the single-site DMFT, a cluster-DMFT scheme is completed by 
self-consistence relation which allows to determine a new Weiss field
through the knowledge of the Green's function.
The two methods we compare in this paper differ in the way this 
self-consistency is implemented.

\subsection{CDMFT method} 
Within CDMFT the cluster self-energy (\ref{sigma}) is used to compute
the $N_c \times N_c$ matrix of ``local'' Green's function
for the cluster, given by 
\begin{equation}
\hat{G}^{loc}(i\omega _{n})=\int_{-\pi /N_{c}}^{\pi /N_{c}}
\frac{1}{(i\omega
_{n}+\mu)\hat{\bf{1}} -\hat{t}_{k}-\hat{\Sigma}^{c}(i\omega _{n})} \frac{dk}{2\pi /N_{c}},
\label{Gloc}
\end{equation}
where the hat labels $N_c\times N_c$ matrices, 
the momentum-integral extends on the reduced Brillouin zone associated
to the $N_{c}$-site cluster, $\hat{t}_{k}$ is the Fourier transform of the
cluster hopping term,\cite{venky} and $\hat{\bf{1}}$ is the $N_c \times N_c$
unit matrix.
$G_{loc}(i\omega _{n})$ and $\Sigma^c(i\omega_n)$ are then used to obtain a 
new Weiss field
\begin{equation}
(\mathcal{G}^{new})_{\mu \nu }^{-1}(i\omega _{n})=\,\Sigma ^{c}_{\mu \nu
}(i\omega _{n})+(G^{loc})^{-1}_{\mu \nu }(i\omega _{n}),  \label{neweiss}
\end{equation}
which determines the new effective action (\ref{seff}), from which a new
cluster Green's function $G_{\mu\nu}(i\omega _{n})$ is obtained and the 
procedure is iterated until
convergence. As noticed above, in this approach the cluster is naturally
taken with open boundary conditions, and breaks the 
full lattice translational invariance. As a consequence, the different sites 
in the cluster are not equivalent, and the local Green's functions on different 
sites may be different.

\subsection{PCDMFT method} 
The lack of translational invariance of the CDMFT may be seen as a limitation
of this method.
The main idea of the PCDMFT approach is therefore to
reintroduce the translational invariance in the CDMFT scheme without
imposing periodic boundary conditions on the cluster, as it is done within
the Dynamical Cluster Approximation.\cite{dca}
The first step of PCDMFT is to 
compute a lattice self-energy with full lattice translational invariance 
by {\it periodizing} the cluster self-energy defined in (\ref{sigma})
\begin{equation}
\Sigma (k,i\omega_n)=\ \frac{1}{N_c}\sum_{\mu \nu }\,e^{ikR_{\mu }}\,\Sigma^{c}_{\mu \nu }(i\omega_n)\,e^{-ikR_{\nu }},  \label{sigmak}
\end{equation}
where $R_{\mu }$ is the position vector of site $\mu $. Notice that the
derivation of the lattice self-energy is not unique, and alternative
estimators have been proposed in Ref. \onlinecite{bk}. The lattice self-energy
naturally defines a translationally invariant lattice Green's function
\begin{equation}
G(k,i\omega _{n})=\,\frac{1}{i\omega _{n}+\mu -t_{k}-
\Sigma (k,i\omega _{n})},
\label{gkappa}
\end{equation}
where $t_k$ is the Fourier transform of the hopping part of the
Hamiltonian.
The key of PCDMFT is to compute the local Green's function for the
cluster degrees of freedom by {\it projecting} back the translationally
invariant Green's function (\ref{gkappa}) onto the cluster sites.
The PCDMFT approximation for the local Green's function of the cluster is
therefore given by 
\begin{equation}
\label{Glocpcdmft}
G^{loc}_{\mu \nu }(i\omega _{n})= \int_{-\pi}^{\pi}
e^{-ikR_{\mu }} G(k,i\omega _{n}) e^{ikR_{\nu }}
\frac{dk}{2\pi}.
\end{equation}
Finally, exactly as in CDMFT, Eq. (\ref{neweiss}) is used to obtain a new 
Weiss field and continue the iterative procedure. 
We emphasize again that the only difference between CDMFT and PCDMFT 
is in the way the local Green's function is computed from the cluster
self-energy, i.e., 
in the difference between Eq. (\ref{Gloc}) and Eq. (\ref{Glocpcdmft}).
The latter equation partially reintroduces the full lattice symmetry
at each iteration.
Finally, it must be noticed that, as shown in Ref.
\onlinecite{bpk}, both the CDMFT\ and PCDMFT methods are  causal, i.e.,
they  produce self-energies with a negative imaginary part.

\subsection{Exact Diagonalization}
Besides the self-consistency defined by the above equations, the CDMFT requires
a solution of the cluster action (\ref{seff}) and the evaluation of the 
Green's function. 
For practical purposes, it is useful to resort to a Hamiltonian formulation,
where the quantum fluctuations on the lattice are realized by hybridization
with a conduction bath. This leads to a cluster-impurity Hamiltonian of the 
form
\begin{eqnarray}
&&\mathbf{H}_{ACI}=\sum_{\mu \nu \sigma }\ E_{\mu \nu}\ c_{\mu
\sigma }^{+}\,c_{\nu \sigma }+
U\ \sum_{\mu }\ n_{\mu \uparrow \ }n_{\mu \downarrow }+ \nonumber \\
&&\sum_{k\sigma }\ \varepsilon \,_{k}\ a_{k\sigma }^{+}a_{k\sigma
}+\sum_{k\mu\sigma}\ (V_{k,\mu}\ a_{k\sigma }^{+}c_{\mu \sigma} + h.c),
\label{aim}
\end{eqnarray}
where the indices $\mu,\nu=1,\ldots,N_c$ label the cluster sites,
$E_{\mu\nu}$ contains the hopping matrix elements inside the
cluster and the chemical potential term. The auxiliary fermionic 
degrees of freedom $a_{k\sigma }$ describe the bath. 
Integrating out the bath degrees of freedom $a_{k\sigma }$ we
obtain an action of the form (\ref{seff}) with
\begin{equation}
\mathcal{G}_{\mu \nu }^{-1}(i\omega_n)=\,i\omega_n\delta _{\mu \nu }-E_{\mu \nu
}-\,\sum_{k}\,\frac{{V_{k,\mu }^{\ast }V_{k,\nu }}}{{i\omega_n
-\,\varepsilon }_{{k}}},  \label{Weiss}
\end{equation}
$\delta_{\mu\nu}$ being the Kronecker delta.

As discussed in Refs. \onlinecite{caffarel,exactdiag} and more
recently in Ref. \onlinecite{potthoff}, to implement the DMFT equations
using an exact diagonalization solver, it is necessary to
discretize the problem by parameterizing the hybridization
function in terms of a finite number of poles. Hence the
hybridization function (last term in  Eq. (\ref{Weiss})) is approximated by
\begin{equation}
\,\sum_{k_i}\,\frac{{V_{k_i,\mu }^{\ast }V_{k_i,\nu
}}}{{i\omega_n -\,\varepsilon _{k_i }}}
\end{equation}
Here the sums run over a discrete set of  $k_i=1,\ldots N_b$.
Notice that in general several $\varepsilon_{k_i}$ can be identical,
i.e., the eigenvalues of the bath can be degenerate. Our numerical
study shows that the self-consistence loop tends to create
degenerate levels. In particular, we have found that each energy
level is typically two-fold degenerate for the two-site cluster
$N_c=2$. This result suggested us to simplify the parameterization
assuming that each level in the bath has such a degeneracy. This
simplified parameterization of the bath, which involves fewer
parameters, typically makes the convergence of (P)CDMFT much
faster (in terms of the number of iterations needed to reach a
given threshold) and allows to easily reach extremely accurate
solution which respect the symmetries of the problem. This
observation also suggests that some values of $N_b$ should
provide worse results than others for the only reason that they
could not allow for the desired degeneracy. As an example, odd
number of bath sites are not the best choice for the two-site
cluster.
For the $N_c=3$ case, we still find that doubly degenerate energy 
levels spontaneously develop in the iterative process, even if the
practical advantage of the degeneracy is much smaller than for $N_c=2$.
Further investigation of larger $N_c$ clusters is needed to fully
understand the origin of the degeneracy.  

We emphasize that the above discussed truncation of
the bath is the only approximation introduced in the exact
diagonalization approach
to (C)DMFT. At each iteration, the Weiss field given by Eq.
(\ref{neweiss}) must be projected on the functional space defined
by (\ref{Weiss}). In practice, the parameters $\varepsilon_k$ and
$V_{k_i}$ which minimize a difference $\Delta$ between (\ref{neweiss})
and (\ref{Weiss}) are used in the next iterations.
What is the optimal criteria for the choice of these poles has
been discussed in the literature, where different criteria have
been used. In this work we describe a simple modification of the
criteria proposed in Ref. \onlinecite{caffarel} which substantially
improves the convergence of the algorithm.
We see that a proper definition of the difference is crucial
as far as the convergence with increasing bath size is concerned.
The substantial independence of the results on $N_b$ (beyond a minimum value) 
observed in Ref. \onlinecite{venky}  is in fact  attributed
to  a definition of functional
distance which weighted too heavily the high frequency region, and we
propose an alternative definition of distance that eliminates
this problem.

\section{results}

The one-dimensional Hubbard model represents an ideal benchmark for our
cluster methods. On the one side we can compare with an exact solution, while on
the other, we know that the single-site DMFT is not able to reproduce
the Mott insulating state for arbitrarily small value of $U$ and the
divergence of the compressibility $\kappa=\partial n/\partial \mu$
approaching the Mott transition.\cite{fisher}
In Ref.  \onlinecite{venky} the ability of CDMFT to reproduce the insulating
state and the dependence of the Mott-Hubbard gap on $U$ has been proven.
Here we focus on the behavior of the density $n$ as a function of the
chemical potential $\mu$ and on the density-driven Mott transition.

\begin{figure}[!htb]
\begin{center}
\includegraphics[width=6cm,height=5cm,angle=-0] {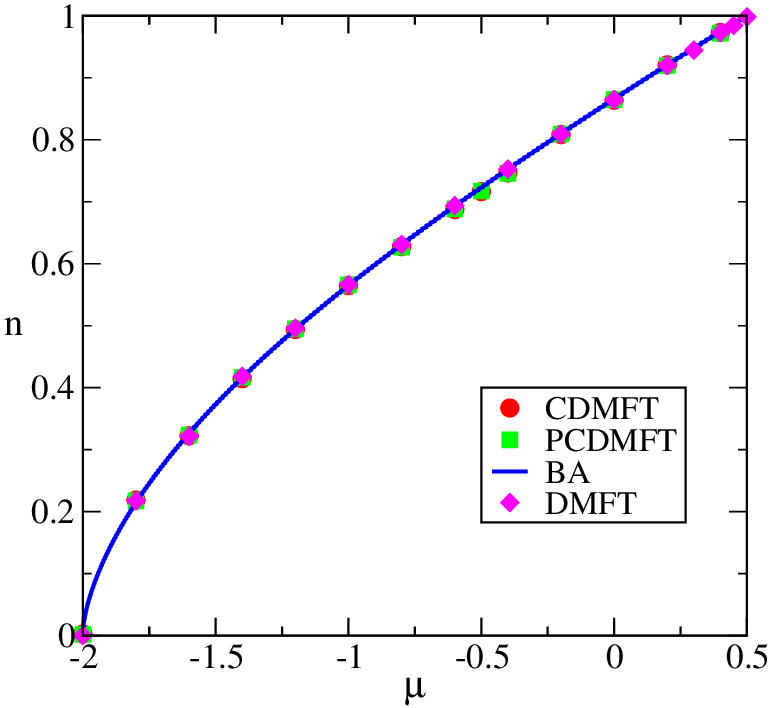}
\end{center}
\caption{Density $n$ as a function of $\protect\mu$ for $U/t=1$ $N_c=2$,$N_b=8$
 within single-site DMFT, CDMFT and PCDMFT compared with the exact solution.}
\label{f1u2}
\end{figure}

\begin{figure}[!htb]
\begin{center}
\includegraphics[width=6cm,height=5cm,angle=-0] {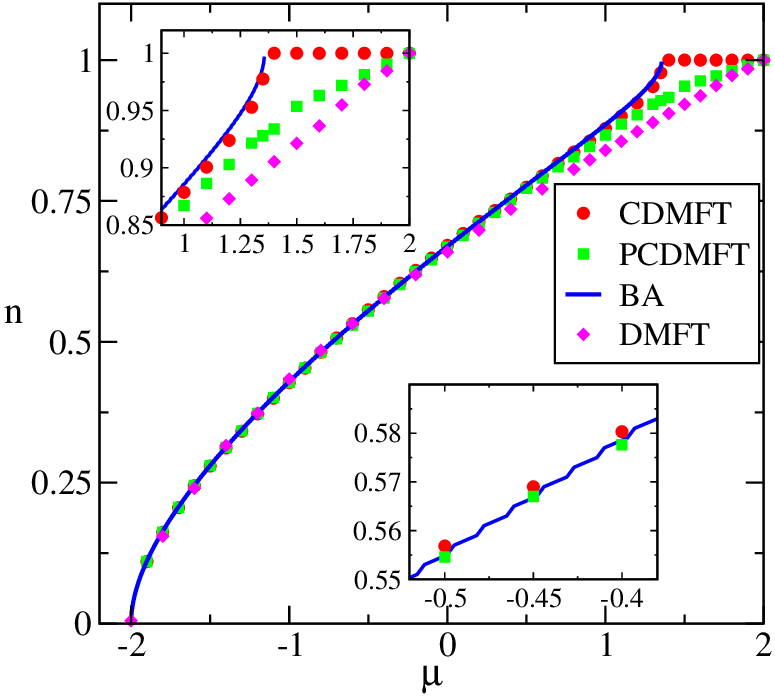}
\end{center}
\caption{Density $n$ as a function of $\protect\mu$ for $U/t=4$, $N_c=2$,$N_b=8
$ . The upper-left inset show a close up of the region close to half-filling,
and the lower-right inset shows a region deep in the metallic regime.}
\label{f2u4}
\end{figure}

\begin{figure}[!htb]
\begin{center}
\includegraphics[width=6cm,height=5cm,angle=-0] {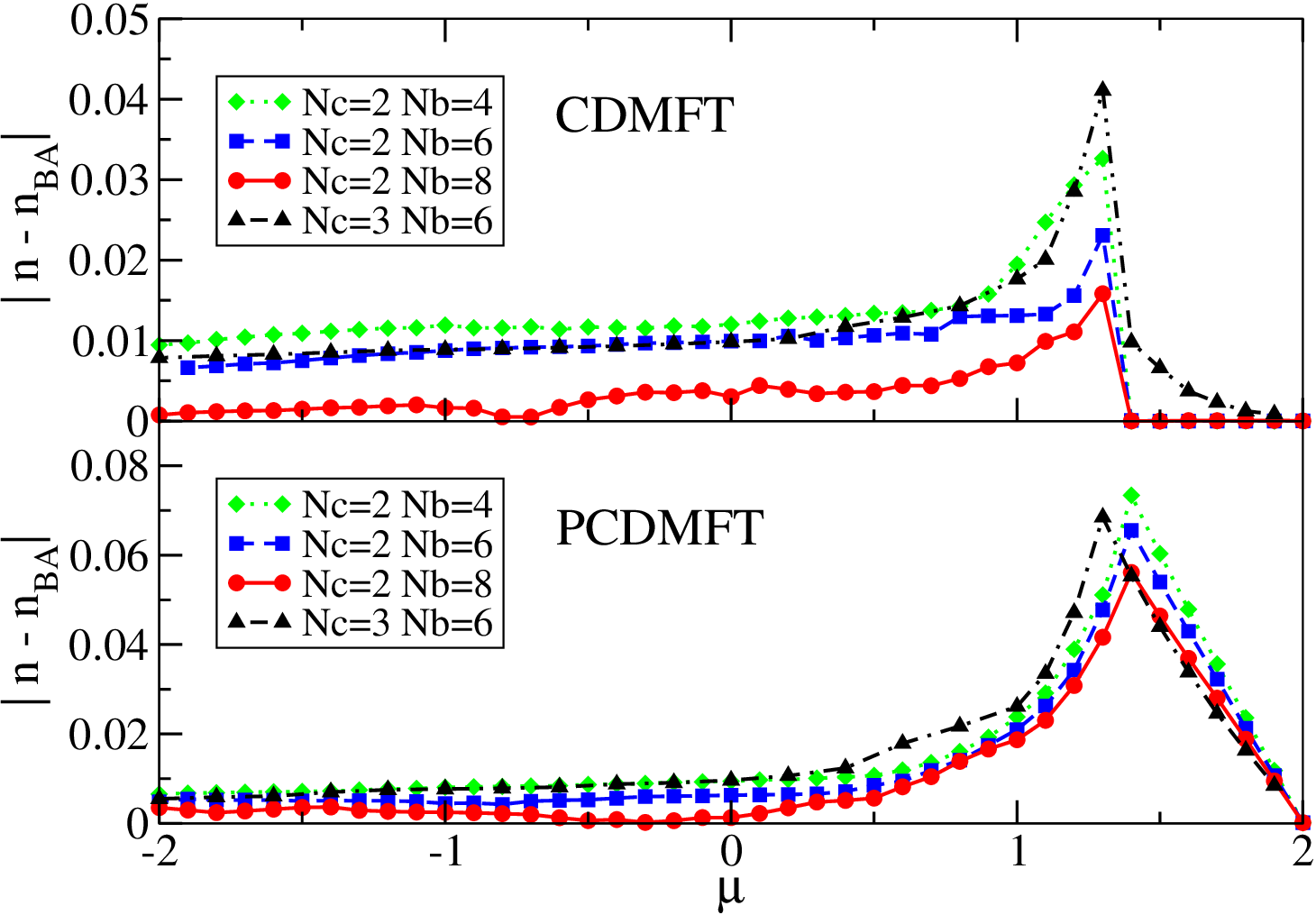}
\end{center}
\caption{Deviations between the CDMFT and PCDMFT densities $n$ with respect
to the exact solution (BA) for $U/t=4$.
In the upper/lower panel we show CDMFT/PCDMFT
for different bath and cluster sizes (see legend).}
\label{f3diff}
\end{figure}

The $n$-$\mu$ curves for $U/t=1$ and $U/t=4$ are reported in Figs. \ref{f1u2}
and \ref{f2u4}, respectively.
The BA results are compared with single-site DMFT and (P)CDMFT for $N_c=2$
and $N_b=8$. In the weakly correlated case ($U/t=1$) the difference between
the different methods are hardly noticeable in the plots, but it must be
noted that the DMFT gives a metallic solution at half-filling as opposed to
the insulating exact solution which both CDMFT and PCDMFT correctly
reproduce already for $N_c=2$. The more correlated $U/t=4$ case emphasizes
the great improvement on DMFT brought by CDMFT. Besides the overall
extremely good agreement with BA, CDMFT is in fact able to closely follow
the compressibility divergence close to the Mott transition, which is missed
by DMFT (upper-left inset of Fig. \ref{f2u4}). PCDMFT is instead unable to
capture this divergence, and gives a finite compressibility, but it becomes
more and more accurate in the extremely doped region ($n \lesssim 0.5$) due to
the explicit reinforcement of lattice periodicity (lower-right inset of
Fig. \ref{f2u4}). The comparison between
the two methods, as well as the dependence of the results on the number of
bath sites is made more quantitative in Fig. \ref{f3diff}, where the
deviation from the exact density is shown for different methods and bath
sizes for $U/t=4$. Not surprisingly, for both techniques, the results
improve as $N_b$ is increased. In the region close to the Mott-Hubbard
transition the CDMFT results nicely approach the exact solution with $\kappa
\to \infty$ as $N_b$ is increased, while for the PCDMFT the improvement with
bath size is not able to reproduce this feature.
Deep in the metallic phase both techniques become extremely accurate, but
the PCDMFT is typically better for $n \lesssim 0.5$.
As an example  for $N_c=2$ and $N_b=8$, the PCDMFT error on the density for 
$\mu=-0.5$ ($n \simeq 0.555$) is $6\times 10^{-4}$, while for CDMFT it is 
$1\times 10^{-3}$ (See the lower-right inset of Fig. \ref{f2u4}).
In both cases $N_b$ small as 6
or 8 is sufficient to achieve remarkable accuracy. We also show results for
$N_c=3$ and $N_b=6$. In this case, even if the cluster size is increased
with respect to $N_c=2$, the
method has some slight problems due to the odd number of sites, which result
in a final critical value of $U/t$ for the Mott transition (even if extremely 
reduced with
respect to DMFT) and a worse agreement with exact results close to the Mott
transition. However, moving to the metallic state, the results rapidly approach
the exact solution as well as the $N_c=2$ ones.

We notice that the expected improvement of results with bath size that we
have found was not observed in Ref. \onlinecite{venky}. We attribute this
difference to the more efficient way to discretize the bath we use in this
work. More precisely, we use a different definition of distance between the
continuous and the discretized Weiss fields that gives more weight to the
low Matsubara frequencies which are important to determine low-energy
physics, namely  $\Delta = \sum_{\nu} \vert \mathcal{G}^{n_s}-\mathcal{G}
\vert / \omega_{\nu}$. We have checked that, using non-weighted differences
and a large number of frequencies, as it was done in Ref. \onlinecite{venky},
the
low-energy part of the spectrum has extremely small weight, and it is
therefore almost irrelevant. In this case the inclusion of further bath 
sites, which would improve the low energy part of the spectrum, is therefore 
basically irrelevant. We notice in passing that the use of the weighted 
distance also favored an easier convergence of the iterations.

These considerations are also supported by the analysis of dynamical
quantities. As in Ref. \onlinecite{venky}, we consider the imaginary
part of the
on-site Green's function $G_{11}$ and the real part of the
nearest-neighbor Green's function $G_{12}$ on the Matsubara axis, which
are plotted for $U/t=$1 and 7 and for $n=1$. single-site DMFT, 
CDMFT and the PCDMFT for $N_c=2$ and $N_b=8$ are compared with the results 
of Density-Matrix Renormalization Group (DMRG), a numerical approach
which is known to provide basically exact results for one-dimensional 
systems (for more details on our calculation of dynamical properties
with DMRG, see Ref. \onlinecite{venky}).
The agreement of both CDMFT and PCDMFT with the virtually exact DMRG
results is extremely good  for $U/t=1$, where the previous CDMFT results
of Ref. \onlinecite{venky} were quite inaccurate.                         
Interestingly, the single-site DMFT completely fails in the description
of dynamical  properties even if the $n$-$\mu$ curve shown in Fig. \ref{f1u2}
is close to the exact solution. In the strong coupling case $U/t=
7$, CDMFT closely follows the DMRG, while PCDMFT is a poorer approximation,
but still substantially better than the single-site DMFT. We
also experienced some problem in obtaining converged PCDMFT solutions for
$U/t > 7$.

\section{conclusions}

We have tested the ability of CDMFT and PCDMFT to capture the
physics of the one-dimensional Hubbard model both in the metallic
and the insulating phase. While we expected cluster methods to
give reasonable results for short-distance high-energy  physical
quantities, we found that both methods are remarkably accurate
for basically all the metallic densities, where the low-energy
physics is important.
The CDMFT method naturally generalizes DMFT to include short-range
dynamical correlations within an open cluster of $N_c$ sites, which
breaks the full lattice translational symmetry. PCDMFT partially restores
the translational symmetry without imposing periodic boundary conditions
to the cluster.
The CDMFT, already for $N_c=2$, is able to reproduce the divergent 
compressibility of the one-dimensional system when the Mott transition
is approached, providing a qualitative
change with respect to single-site DMFT.
The PCDMFT, due to the restored translational invariance, gives slightly 
better results than CDMFT deep in the metallic regime (far from the 
Mott transition), where correlation effects are smaller, but fails in 
reproducing correctly the physics close to the Mott transition.
The remarkably good results
of CDMFT (either from the metallic to the insulating state) and of
PCDMFT (in the metallic state only) are also due to a technical
improvement in the truncation procedure of the bath hybridization 
function of the impurity problem. A systematic improvement of 
the accuracy of the results by
increasing the bath size is observed. We also find that some
degeneracy of the bath energy levels automatically develops if
not explicitly imposed. Explicit enforcement of degenerate levels
with the correct degeneracy strongly favors convergence and
accuracy of the solutions. These methodological advances will be
useful in further studies of more complex systems.

M. Capone acknowledges the warm hospitality of Rutgers University,
where this work has been carried out, and the support from the
Physics Department of the University of Rome "La Sapienza" and 
INFM UdR Roma1. This work was supported by
the NSF under grant DMR-0096462, the  Italian Miur 
Cofin 2003 and INFM.

\begin{figure}[!htb]
\begin{center}
\includegraphics[width=6cm,height=5cm,angle=-0] {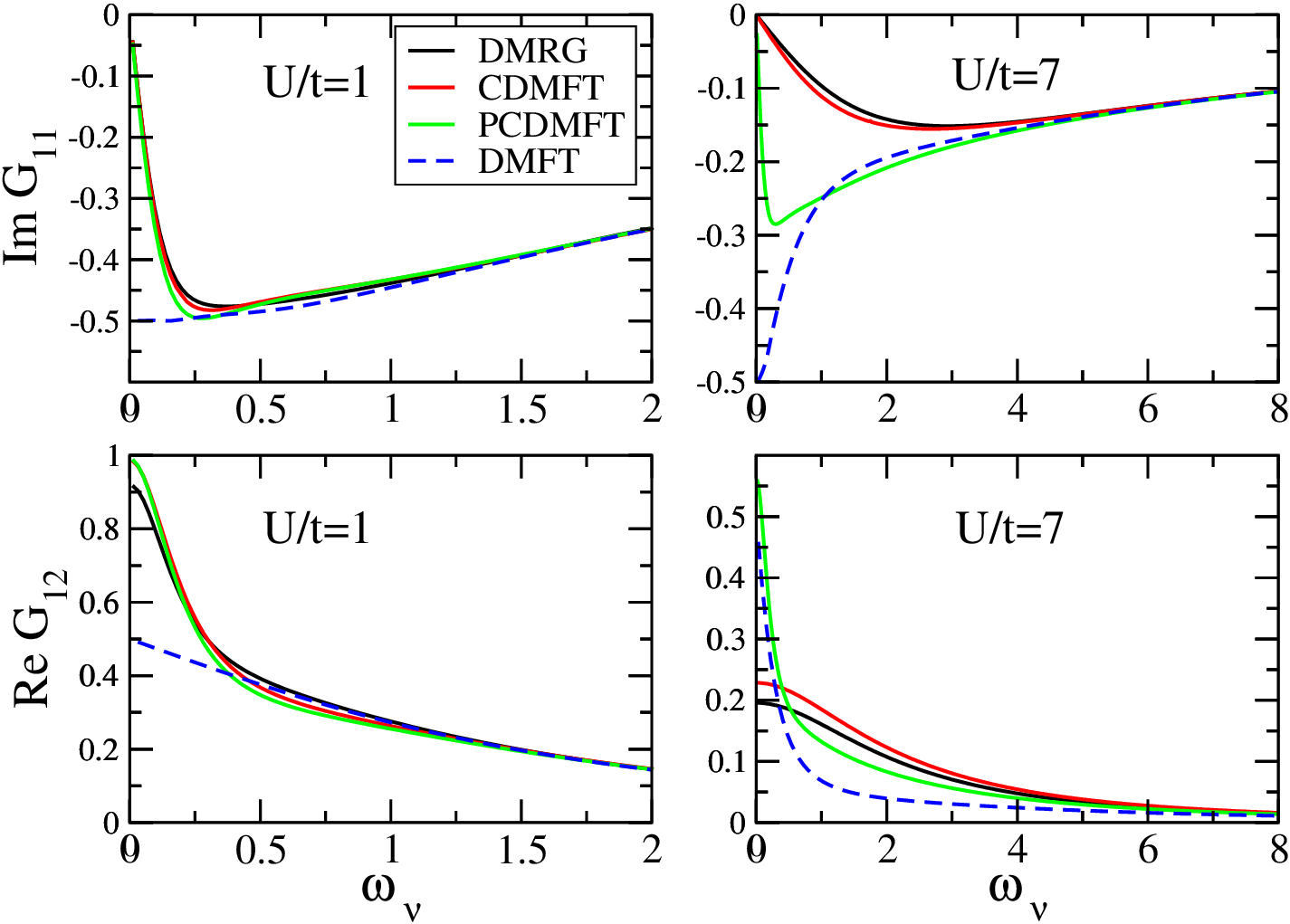}
\end{center}
\caption{Imaginary part of the local Green's function $G_{11}$ and real part
of nearest-neighbors Green's functions $G_{12}$ for $n=1$.
Single-site DMFT and two-site CDMFT
and PCDMFT are
compared with DMRG for $U/t=1$ (left) and $U/t=7$ (right).}
\label{f5dynamical}
\end{figure}

\end{document}